\documentclass{PoS}

\title{The Detailed Chemical Abundance Patterns of M31 Globular Clusters}

\ShortTitle{The Detailed Chemical Abundance Patterns of M31 Globular Clusters}

\author{\speaker{Janet E. Colucci}
\thanks{The data presented herein were obtained at the W. M. Keck Observatory, which is operated as a scientific partnership among the California Institute of Technology, the University of California and the National Aeronautics and Space Administration. The Observatory was made possible by the generous financial support of the W. M. Keck Foundation.}
\\
        University of California Santa Cruz\\
        E-mail: \email{jcolucci@ucolick.org}}

\author{Rebecca A. Bernstein \\
        University of California Santa Cruz\\
        E-mail: \email{rab@ucolick.org}}

\author{Judith Cohen \\
     California Institute of Technology \\
        E-mail: \email{jlc@astro.caltech.edu}}

\abstract{We present detailed chemical abundances for $>$20 elements in $\sim$30 globular clusters in M31. These results have been obtained using high resolution ($\lambda/\Delta\lambda\sim$24,000) spectra of their  integrated light and analyzed using our original method. The globular clusters have galactocentric radii between 2.5 kpc and 117 kpc, and therefore provide abundance patterns for different phases of galaxy formation recorded in the inner and outer halo of M31. We find that the clusters in our survey have a range in metallicity of  $-2.2<$[Fe/H]$<-0.11$. The inner halo clusters cover this full range, while the outer halo globular clusters at R$>$20 kpc have a small range in abundance of [Fe/H]$=-1.6 \pm 0.10$.  We also measure abundances of alpha, r- and s-process elements. These results constitute the first abundance pattern constraints for old populations in M31 that are comparable to those known for the Milky Way halo.}

\FullConference{XII International Symposium on Nuclei in the Cosmos,\\
		August 5-12, 2012\\
		Cairns, Australia}

\begin{document}

\section{Introduction}
    \vspace{-0.3cm}

Study of M31, the Milky Way's nearest massive neighbor, is interesting for many reasons. One of the most fundamental questions is whether the characteristics of the  M31 spiral galaxy support the assertion that the Milky Way  is a ``normal'' spiral galaxy.   This is important because we can study the properties of the Milky Way in great detail, and studies of our own galaxy by necessity are the foundation for our understanding of how galaxies in general form and evolve.   As the next closest massive galaxy, M31 is the first place to test galaxy formation theories developed from studies of the Milky Way, and in some respects is a more ideal test-case because M31 can be observed as a whole from the outside, whereas study of our own galaxy is complicated by our position within it.  However, the distance to M31 means that we are unable to study its individual stars at the same level of detail that we can obtain in the Milky Way, using high resolution high signal-to-noise ratio (SNR) spectra and detailed chemical abundance analyses.

With the development of our original technique for abundance analysis of high resolution integrated light (IL) spectra of globular clusters (GCs), we can now make significant advances in chemical evolution studies of distant massive galaxies.  Unresolved GCs, which are luminous and therefore observationally accessible to large distances, can be used to learn about the chemical enrichment and formation history of other galaxies, just as they were originally used to learn about the formation of the Milky Way \cite{1}. 
 Our technique has been developed and demonstrated on resolved GCs in the Milky Way and Large Magellanic Cloud (LMC) in a series of papers \cite{2,3,4,5,6, 7}.  These works demonstrate that the IL analysis provides accurate Fe abundances and [X/Fe] ratios to $\sim$0.1 dex, as well as  distinguishes ages  for GCs with a range in properties, including [Fe/H] of $-2$ to $+0$ and ages from 0.05 to 12 Gyr.

With this method, we have now  begun an unprecedented study of the chemical composition of the GC system of M31.  Detailed abundances of $\sim20$ elements were presented for a  pilot sample of 5 M31 GCs in \cite{4}.  As part of this ongoing project, here we  extend the sample of \cite{4} and present ages, Fe, $\alpha$-element and neutron capture abundances of an additional 22 GCs in M31. 		
		
\section{Observations}
    \vspace{-0.3cm}

We obtained high resolution IL spectra of the M31
GCs using the  HIRES spectrograph
  on the Keck I telescope. The data were taken over several observing runs from 2006-2011.  In all observing runs we used  identical setups that utilized  the D3 decker, which has a  slit size of $1.7"\times7.0"$ and spectral resolution of $R=24,000$, which is sufficient to resolve individual spectral lines of GCs with velocity dispersions,  $\sigma_{v}>\sim7$kms$^{-1}$. 
 The wavelength coverage  is approximately
3800$-$8300 \AA.  Exposure times were between 1$-$5 hours for each GC.
 The SNR at  6000 \AA~ is approximately 60-90.  Data were reduced with standard routines in the
HIRedux pipeline.\footnote{ http://www.ucolick.org/~xavier/IDL/index.html}

\section{Analysis}
    \vspace{-0.3cm}

As in all of our  our previous spectroscopic analyses we first measure individual absorption line equivalent widths (EWs) using the semi-automated program GETJOB \cite{8}. 
 Line lists and oscillator strengths were
taken from \cite{3,4,6} and references therein.  
   In this work, we also use our IL spectral synthesis code \cite{7} for the best possible accuracy in our final measurements for Ca, Si, Eu, Ba and Y.  Abundances for Ba and Eu include corrections for hyperfine splitting, as described in \cite{7}.

The IL abundance analysis is described in detail in \cite{3,4,5}.  In this analysis  we create synthetic GC color magnitude diagrams (CMDs)  using the Teramo isochrones \cite{9}  
 without convective overshooting,  with extended asymptotic giant
branch (AGB), 
and mass$-$loss parameter of $\eta$=0.2.  More discussion about the testing and choice of appropriate isochrones can be found in \cite{3,5,6}.
 Flux-weighted synthesized EWs of lines are calculated using our
routine ILABUNDS \cite{3}, which utilizes 
spectral synthesis routines from the 2010 version of MOOG \cite{10}.
We use the  ODFNEW and AODFNEW  model stellar atmospheres from Kurucz.\footnote{http://kurucz.harvard.edu/grids.html} 
All abundances are calculated under the assumption of local
thermodynamic equilibrium (LTE). 

The age and [Fe/H]  solutions for  each cluster are  identified as the range in synthetic CMD ages and [Fe/H] that produce the most self-consistent results using the 10$-$80 individual Fe I lines measured in each cluster.  
The best solutions have the smallest statistical error ($\sigma_{{\rm N}}$), and minimal dependence of Fe I abundance with line excitation potential (EP), wavelength, and EW.  For each cluster there is a range in CMD ages that produce similarly self-consistent solutions.  For older clusters this range is typically 10$-$15 Gyr, and leads to  an internal  systematic uncertainty in [Fe/H] of $\lesssim$ 0.05 dex.   For the total uncertainty in [Fe/H] for each cluster, we add this internal  systematic age uncertainty, $\sigma_{\rm {Age}}$, in quadrature with the
statistical error in the mean abundance, $\sigma_{ {\rm N}} / \sqrt{{\rm N} -1} $, where N is the number of lines measured for each element.
The best CMD solution is used to calculate abundances for the other elements.

\begin{figure}
    \vspace{-0.9cm}
  \includegraphics[angle=90,width=0.45\textwidth]{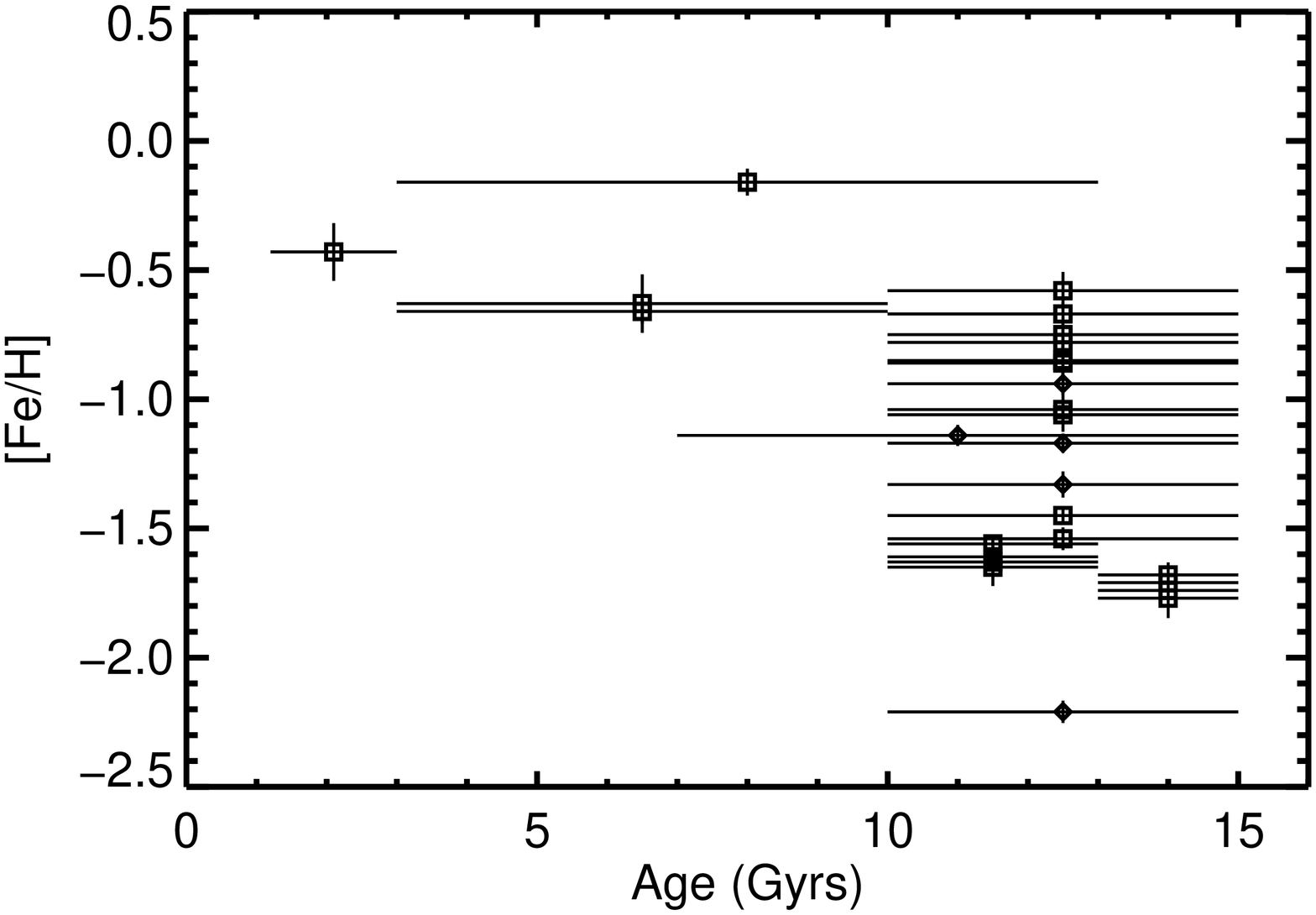}
\hspace{0.04\textwidth}
  \includegraphics[angle=90,width=0.45\textwidth]{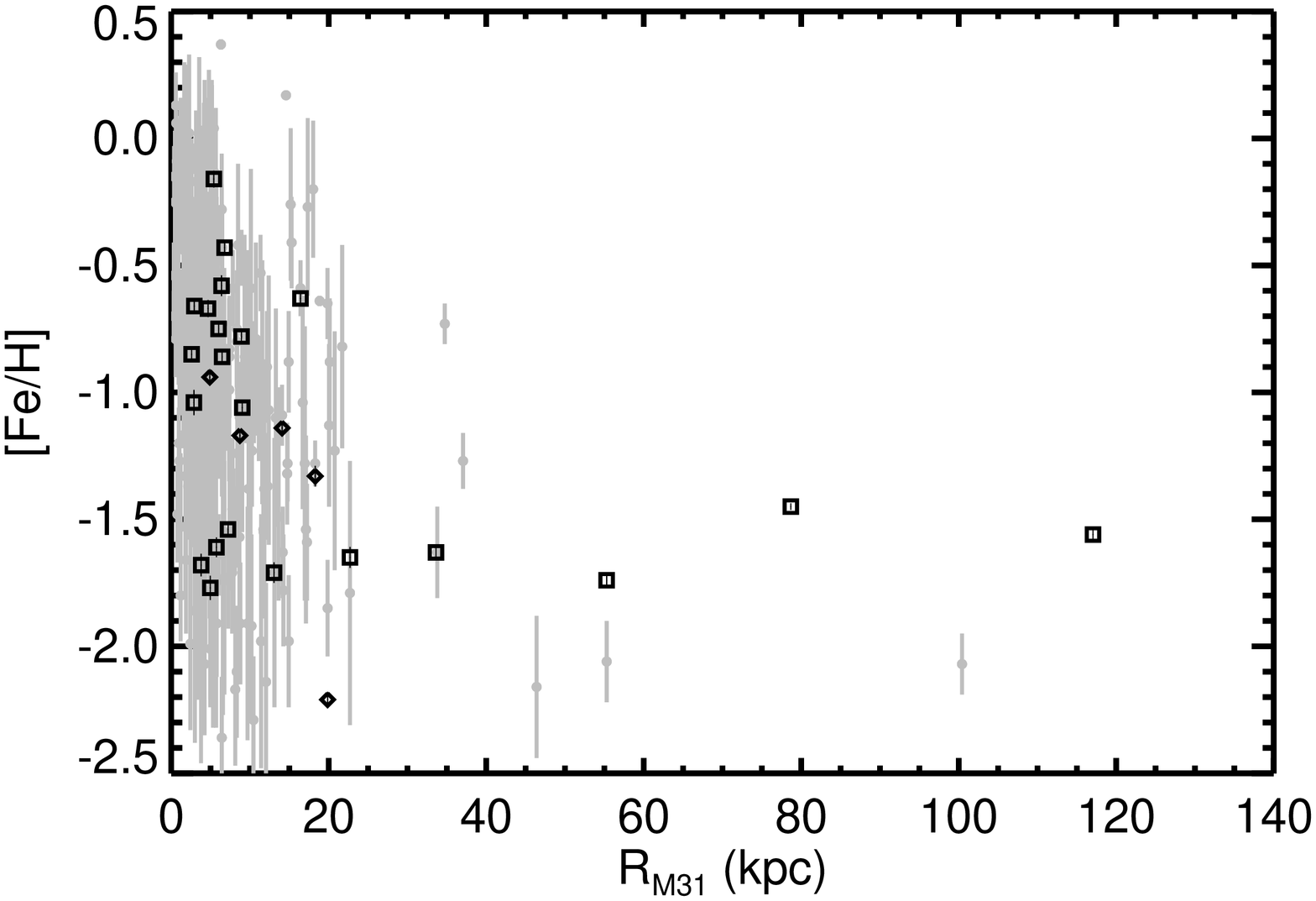}
    \vspace{-0.5cm}
\caption{{\bf Left:} The age-metallicity relationship. Black diamonds are from \cite{4}, and black squares are from this work. {\bf Right:} The behavior of [Fe/H] with projected galactocentric radius from M31, where radii are taken from \cite{11}. Gray points are from \cite{12}, and measured using Lick system line indexes.}
\label{fig:feh}
\end{figure}

\section{Results and Conclusions}
    \vspace{-0.3cm}

In Figure \ref{fig:feh} we show the age-metallicity relationship for the 22 M31 GCs analyzed in this work and the 5 GCs analyzed previously \cite{4}.  
Figure \ref{fig:feh} also shows the M31 GC [Fe/H] results against M31 projected galactocentric radii from  \cite{11}.  The low resolution results of \cite{12} are also shown for comparison and to highlight the greater precision of our technique when compared to low resolution results. We find a spread in [Fe/H] for GCs within R$_{\rm M31}$$\sim$20 kpc, and a fairly constant [Fe/H] for GCs at R$_{\rm M31}$$>$20 kpc.  The mean value for the 5 GCs in our sample that lie at R$_{\rm M31}$$>$20 kpc is [Fe/H]$=-1.63 \pm 0.10$.  We therefore confirm the previous results of \cite{13,14}, who also found a nearly constant [Fe/H] for outer halo GCs. 
Like our analysis, \cite{14} found  a flat metallicity distribution for GCs outside 30 kpc, with a mean metallicity  of  [Fe/H]=$-1.6$, which was estimated   from line indexes. 
We also note that at least one GC in our sample has an intermediate age (1-3 Gyr) at high [Fe/H];  this cluster has disk-like kinematics \cite{15} and lies within 7 kpc.

\begin{figure}
    \vspace{-1.1cm}
  \includegraphics[width=0.48\textwidth]{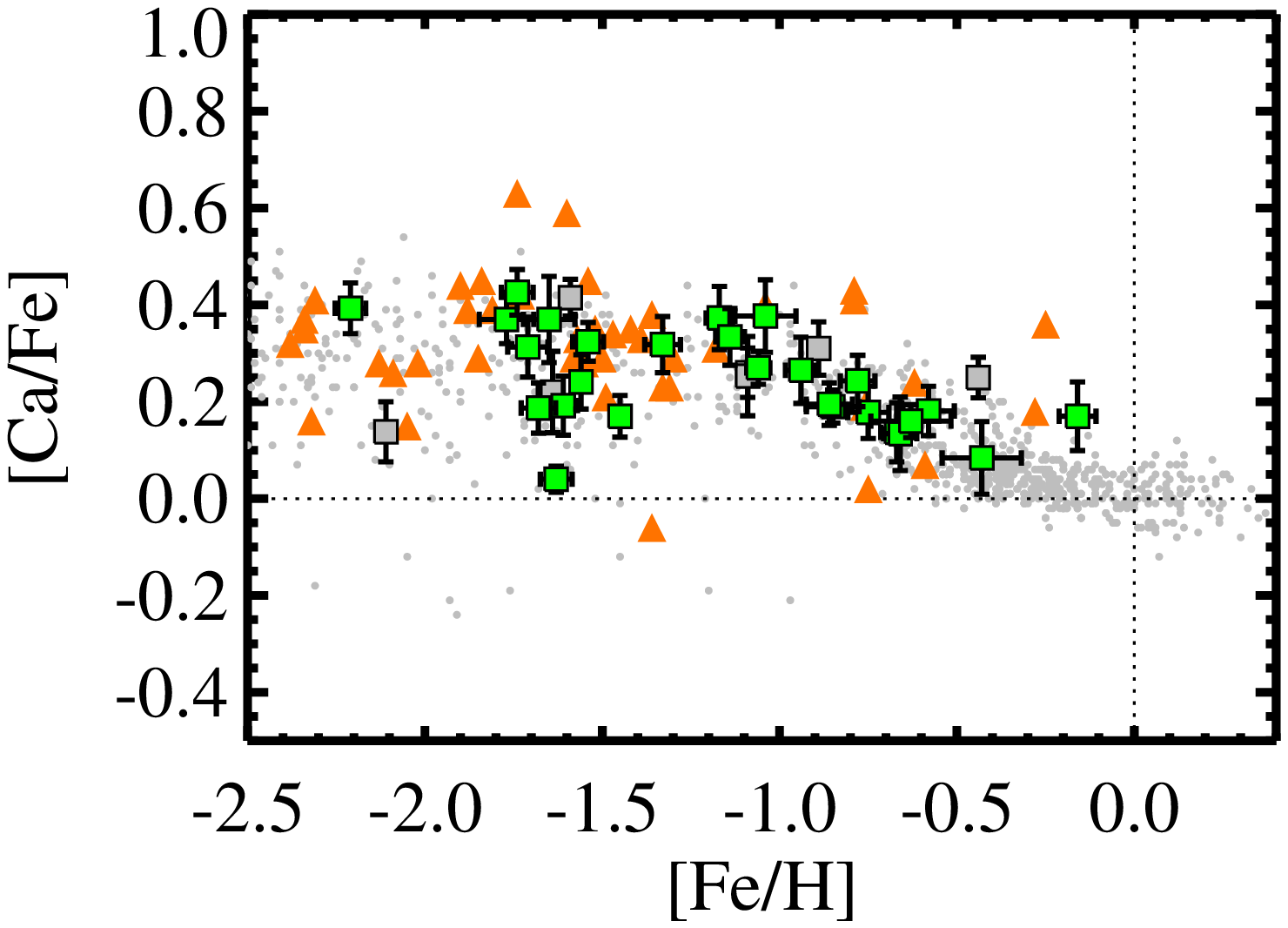}
\hspace{0.04\textwidth}
  \includegraphics[width=0.48\textwidth]{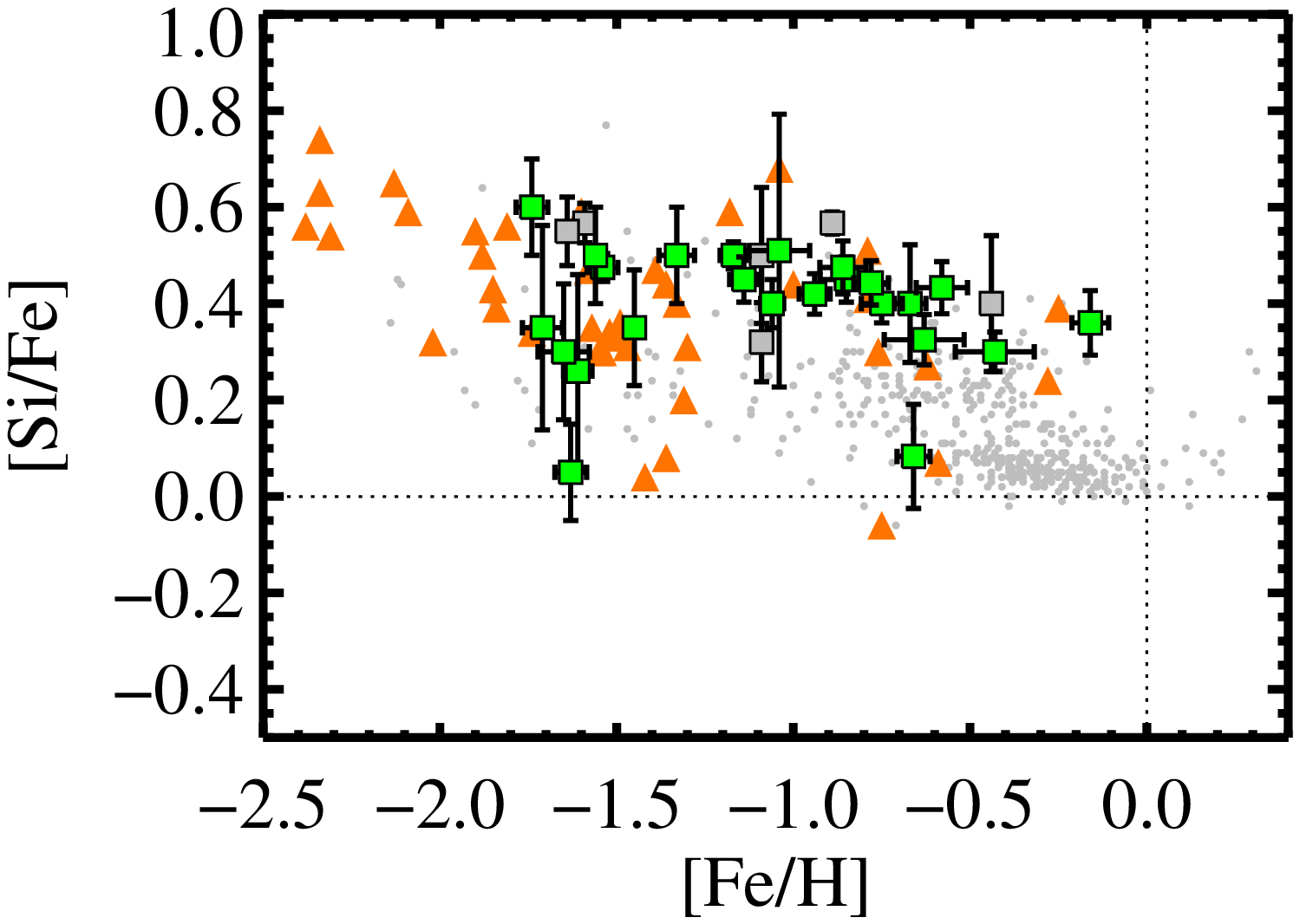}
  \vspace{-1.cm}
\caption{$\alpha$-element abundances in our sample to date.  Green squares show M31 GCs, Grey squares show MW IL abundances from our training set GCs \cite{5}.  Small gray circles and orange triangles, respectively, show  MW field stars \cite{16} and MW GC stars \cite{17}. }
\label{fig:alpha}
\end{figure}

\begin{figure}
    \vspace{-0.5cm}
  \includegraphics[width=0.48\textwidth]{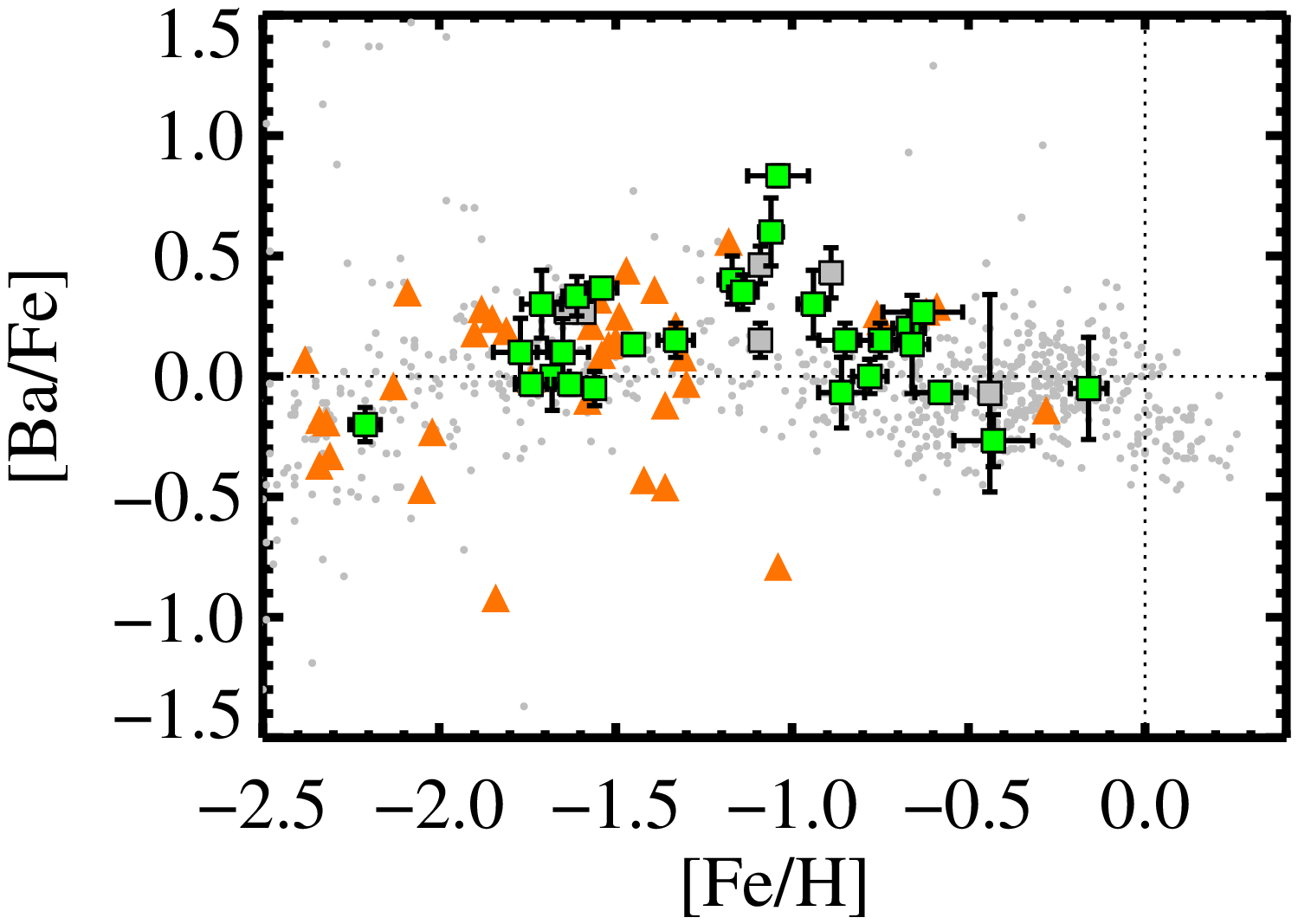}
\hspace{0.04\textwidth}
  \includegraphics[width=0.48\textwidth]{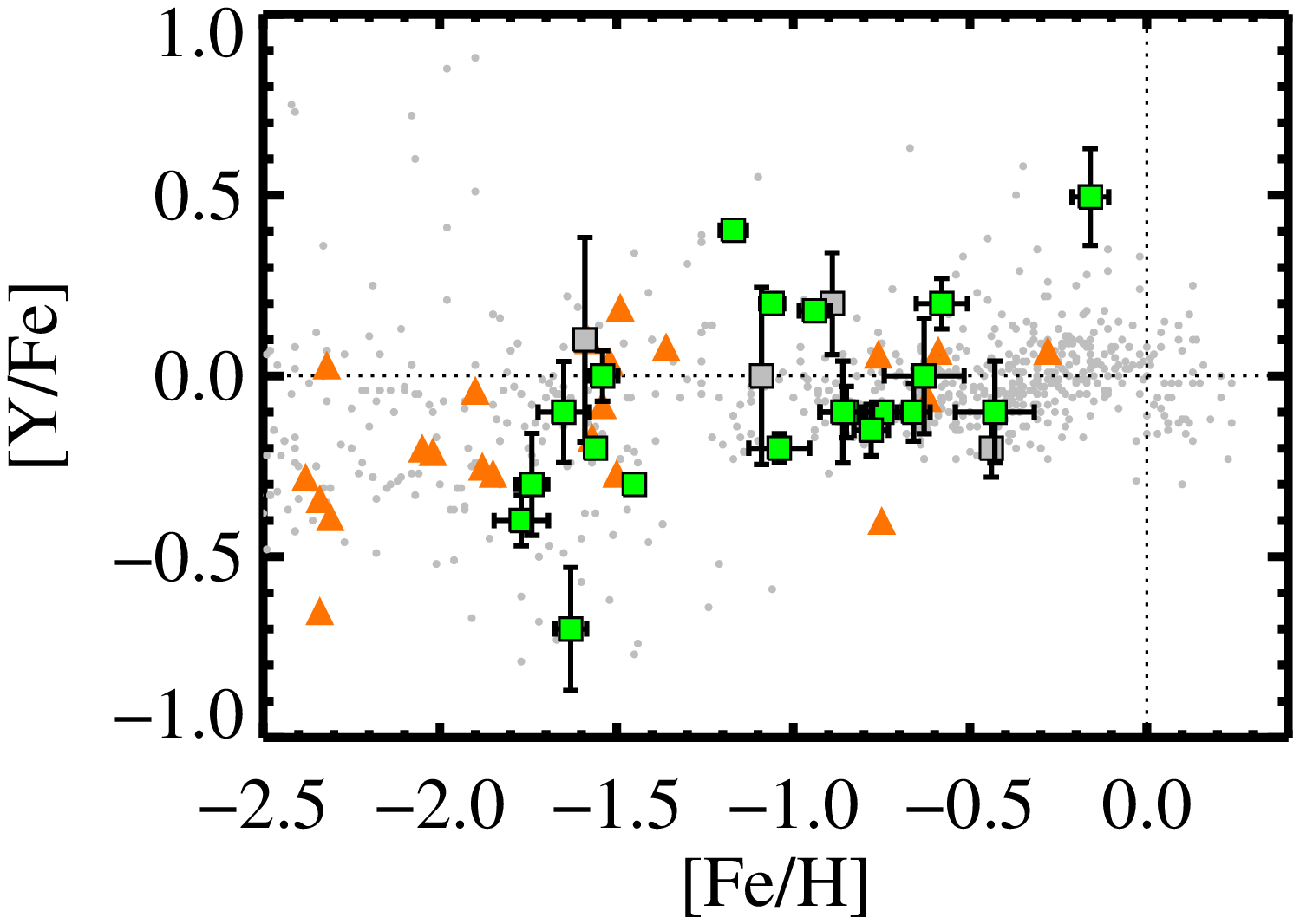}
    \vspace{-1.cm}
\caption{[Ba/Fe] and [Y/Fe], symbols are the same as in Figure 2. }
\label{fig:bay}
\end{figure}

\begin{figure}
    \vspace{-0.5cm}
  \includegraphics[width=0.48\textwidth]{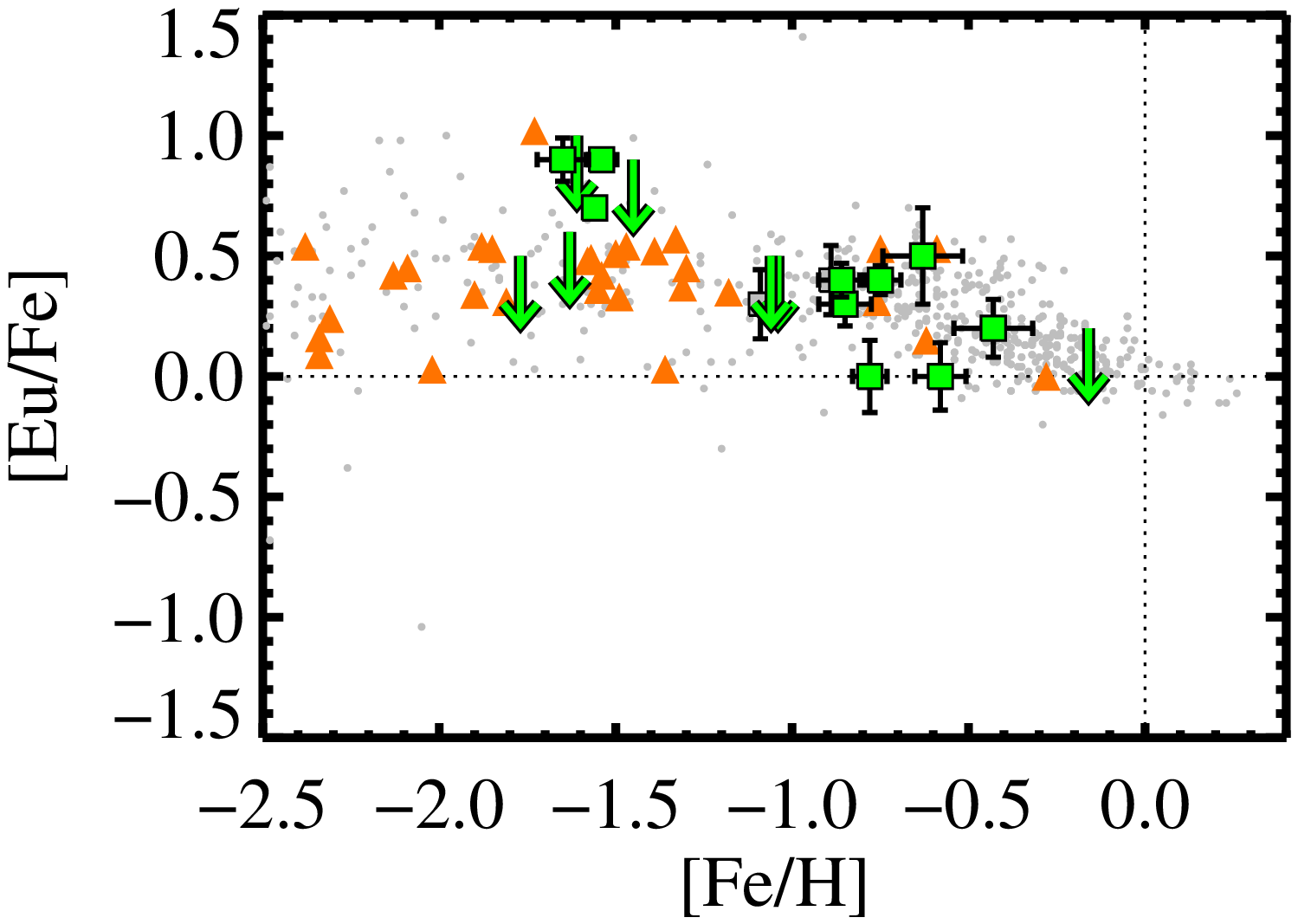}
\hspace{0.04\textwidth}
  \includegraphics[width=0.48\textwidth]{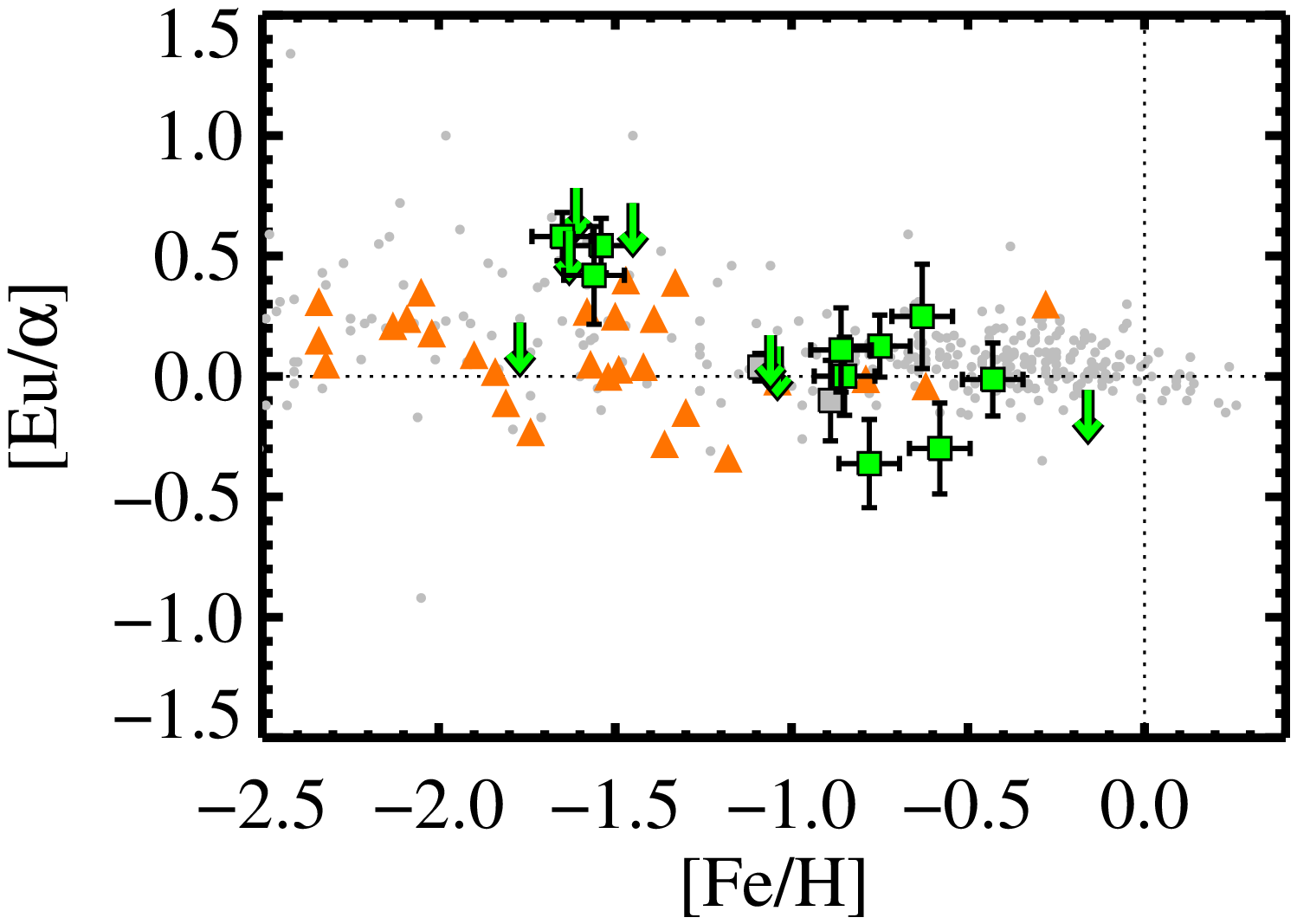}
    \vspace{-1.cm}
\caption{ [Eu/Fe]  and [Eu/$\alpha$],  symbols are the same as in Figure 2.}
\label{fig:eu}
\end{figure}

\begin{figure}
    \vspace{-1.1cm}
  \includegraphics[width=0.48\textwidth]{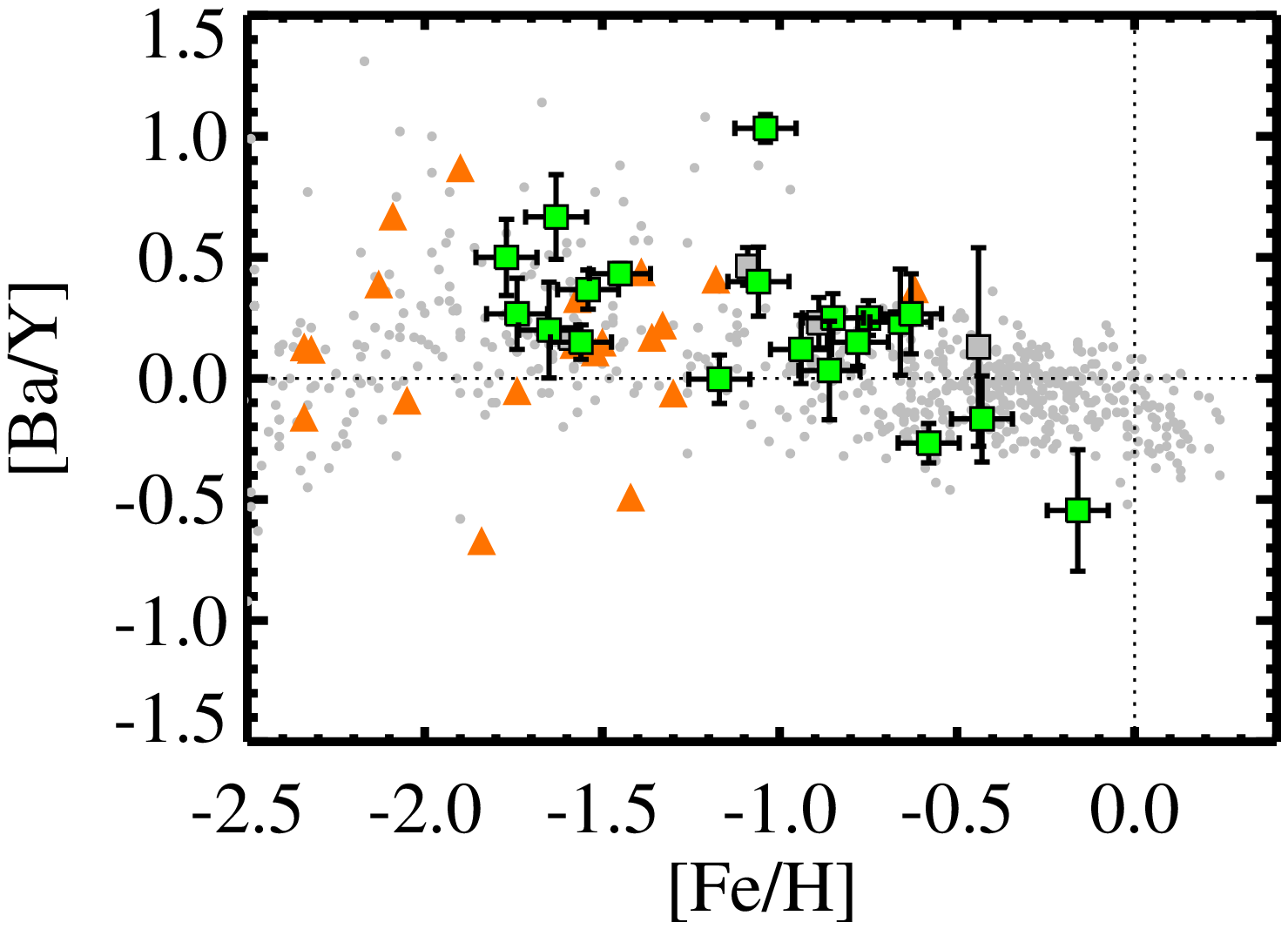}
\hspace{0.04\textwidth}
  \includegraphics[width=0.48\textwidth]{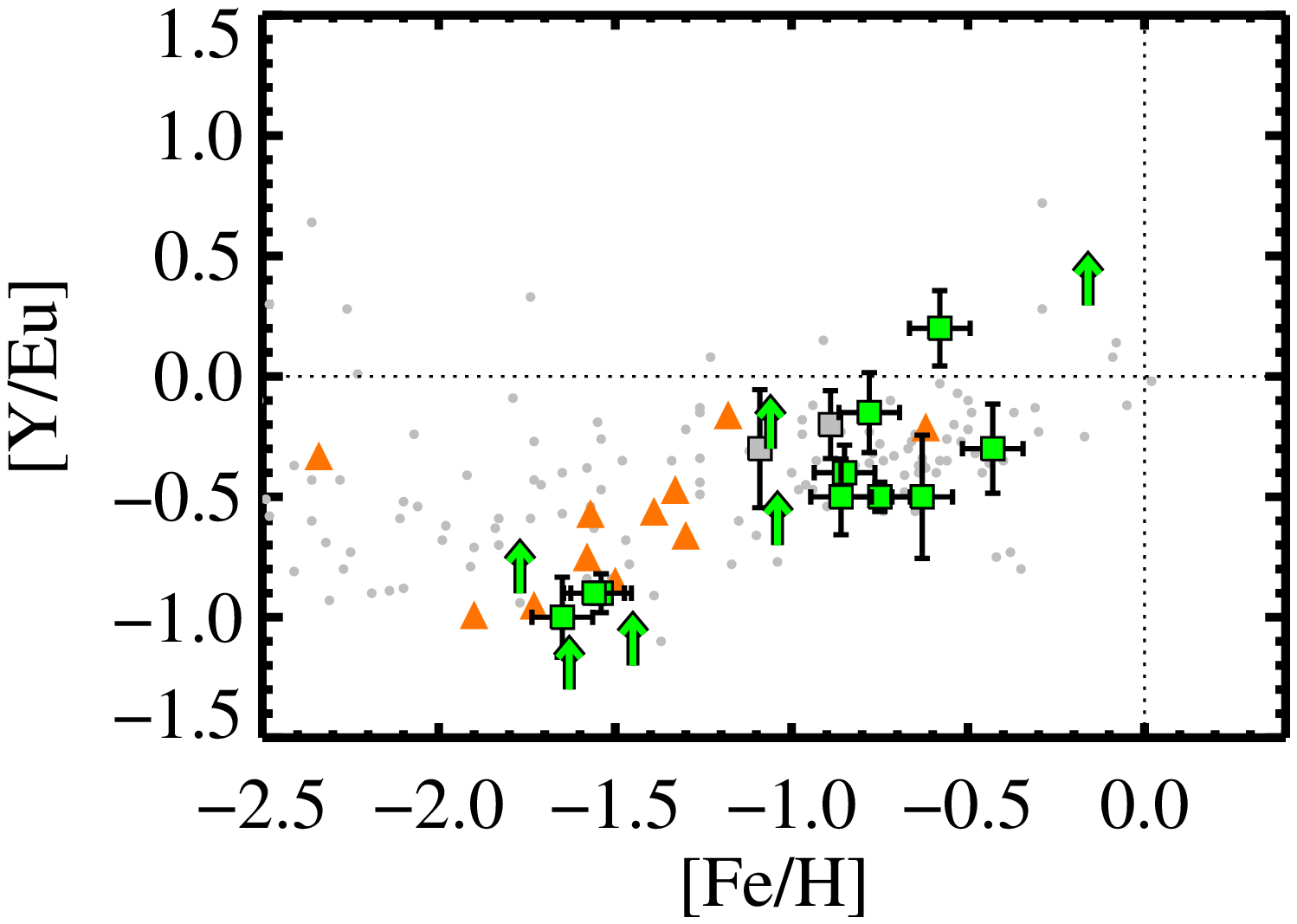}
    \vspace{-1.cm}
\caption{ [Ba/Y] and [Y/Eu],  symbols are the same as in Figure 2. }
\label{fig:yeu}
\end{figure}

In Figures \ref{fig:alpha}--\ref{fig:eu}, we show the [X/Fe] as a function of [Fe/H] for Ca, Si, Ba, Y, and Eu.  For the $\alpha$-elements Ca and Si, we find a plateau value that is similar to the value in MW GCs and field stars from \cite{16,17}, which is consistent with rapid early star formation in M31. Note that our  results do not confirm the  line index results from \cite{18,19}.  The underestimations of [$\alpha$/Fe] in \cite{18,19} are likely due to Mg depletion within some of the cluster stars; we first published evidence of this effect being visible in the integrated light of GCs in  \cite{4,7}.  The  star-to-star variations in the light elements Na, O, Mg, and Al, are a well known phenomenon, and discussed in the review by \cite{20}.  Most recently \cite{21} confirmed, using a slightly  different IL analysis,  that star-to-star variations in Mg  are detectable in IL, and therefore  Mg is not a good proxy for [$\alpha$/Fe] in IL GC spectra.      In Figure \ref{fig:alpha} there is  a clear knee visible in the [Ca/Fe] values, which closely matches the abundance patterns of MW field stars.  Obvious outliers  in the abundances of our GC sample include at least one GC with high [Fe/H] and high [$\alpha$/Fe], and one GC with low [Fe/H] and low [$\alpha$/Fe], the latter of which may be associated with an accretion event \cite{22}.   

Note that the abundances shown in Figures 3 and 4 are the first results for r- and s-process abundances in old populations in M31. In general, we find that the Ba, Y, and Eu are consistent with MW heavy element abundances. The [Eu/Fe] values for GCs with [Fe/H]$<-1.5$ show some suggestion of being higher than the mean of MW GCs, although they are still within the range exhibited by the MW. A larger sample of low metallicity Eu abundances is necessary to confirm this result.

In Figure \ref{fig:eu} - \ref{fig:yeu} we show key neutron capture abundance ratios as a function of overall metallicity. These are particularly important diagnostics as [Ba/Y] identifies the yield of second peak to first peak s-process elements, while [Y/Eu] identifies the yield of light s-process to r-process elements. [Eu/$\alpha$] is a diagnostic for different elements that are primarily produced in Type II supernovae.   We find that this sample of M31 GCs is dominated by r-process enrichment at low metallicity, and the s-process influence appears to increase around [Fe/H]$\sim -1$, similar to the MW. These results, like the [$\alpha$/Fe] results,  are consistent with rapid, early star formation in M31.

In summary, we present ages and abundances of Fe, Ca, Si, Ba, Y, and Eu for a total sample of 27 GCs in M31.  For the abundances shown here, M31 generally looks like the MW; even this basic similarity has been controversial to assert from previous line index analyses.   Our analysis of the IL spectra of extragalactic GCs demonstrates the importance of quantitative, accurate abundance ratios in clarifying the chemical enrichment and formation history of galaxies beyond the Milky Way.  M31 is the first of many galaxies within the Local Group and nearby Groups that we are studying with this technique.

\end{document}